\documentclass[lettersize,journal]{IEEEtran}

\usepackage{cite}
\usepackage{amsmath,amssymb,amsfonts}
\usepackage{graphicx}
\usepackage{textcomp}
\usepackage[ruled,linesnumbered,vlined]{algorithm2e}
\usepackage{stfloats}
\usepackage[hyphens]{url}
\usepackage{verbatim}
\usepackage{dsfont}
\usepackage{algpseudocode}
\usepackage{pifont}
\usepackage{color}
\usepackage{soul}
\usepackage{booktabs}
\usepackage{tabularx}
\usepackage{flushend}
\usepackage[normalem]{ulem}
\usepackage[table,xcdraw]{xcolor}
\usepackage{array}
\usepackage{threeparttable}

\usepackage[colorlinks=true,
            urlcolor=black,
            linkcolor=red,
            anchorcolor=black,
            citecolor=green]{hyperref}

\usepackage{makecell}
\usepackage{multirow}
\usepackage{nicematrix}
\usepackage{float}
\usepackage[numbers,sort&compress]{natbib}
\usepackage[justification=justified]{caption}
\usepackage{subfigure}

\begin{document}

\title{Enhancing LUT-based Deep Neural Networks Inference through Architecture and Connectivity Optimization}

\author{
        {Binglei Lou, Ruilin Wu, Philip Leong}\\
        {School of Electrical and Computer Engineering}\\
       {The University of Sydney, 2006, Australia}\\
       {Email: \{binglei.lou,ruilin.wu,philip.leong\}@sydney.edu.au}       
}

\maketitle

\begin{abstract}
Deploying deep neural networks (DNNs) on resource-constrained edge devices such as FPGAs requires a careful balance among latency, power, and hardware resource usage, while maintaining high accuracy. Existing Lookup Table (LUT)-based DNNs---such as LogicNets, PolyLUT, and NeuraLUT---face two critical challenges: the exponential growth of LUT size and inefficient random sparse connectivity.

This paper presents SparseLUT, a comprehensive framework that addresses these challenges through two orthogonal optimizations. First, we propose an architectural enhancement that aggregates multiple PolyLUT sub-neurons via an adder, significantly reducing LUT consumption by 2.0$\times$–13.9$\times$ and lowering inference latency by 1.2$\times$–1.6$\times$, all while maintaining comparable accuracy. Building upon this foundation, we further introduce a non-greedy training algorithm that optimizes neuron connectivity by selectively pruning less significant inputs and strategically regrowing more effective ones. This training optimization, which incurs no additional area and latency overhead, delivers consistent accuracy improvements across benchmarks---achieving up to a 2.13\% gain on MNIST and 0.94\% on Jet Substructure Classification compared to existing LUT-DNN approaches.
\end{abstract}


\section{Introduction}
\IEEEPARstart{D}{eep} neural networks (DNNs) have significantly improved our ability to solve pattern recognition problems across diverse data formats, including images, video, speech, audio and text~\cite{lecun2015deep}. Field-Programmable Gate Arrays (FPGAs) provide a unique implementation platform for deploying DNNs as they enable better integration with other system components such as networks and video decoders, allow signal processing to be combined with DNNs, and can operate with lower energy and latency than graphics processing units, particularly in real-time inference tasks.

\begin{figure}[b]
    \centerline{\includegraphics[width=0.4\linewidth]{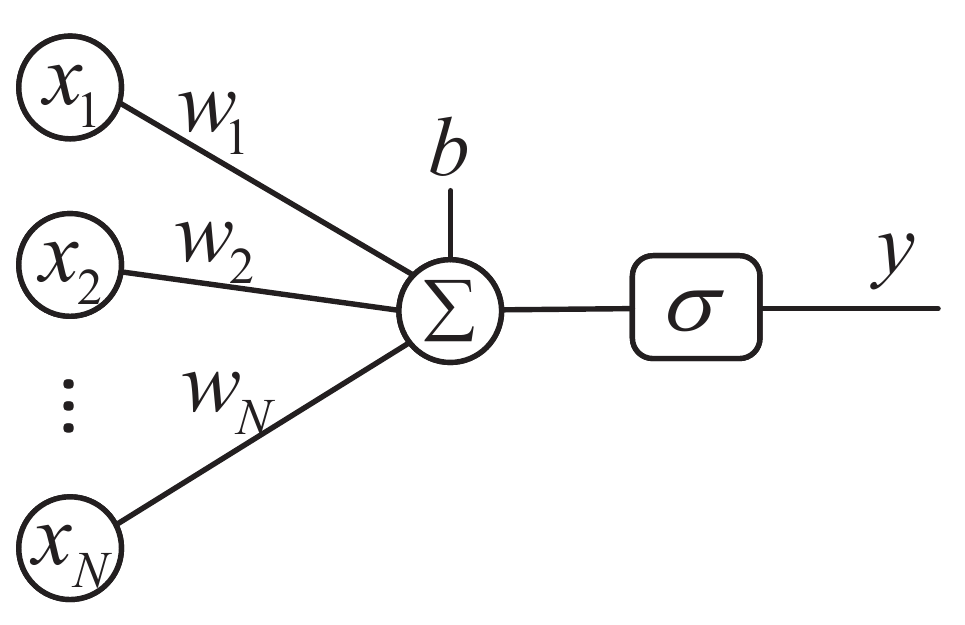}}
    \caption{Neuron computation with fan-in $N$. LUT-DNNs typically select a random subset (size $F \ll N$) of the inputs. SparseLUT is a training scheme to select the inputs while maximizing accuracy.} 
    \label{fg:neuron}
\end{figure}

In an artificial neural network, the output of a neuron is given by 

\begin{equation}
y = \sigma \left( {\sum\nolimits_{k = 1}^{N} {{w_k}{x_k} + b} } \right)
\label{eq:neuron}
\end{equation}

where $N$ is the number of inputs, $x_i$ are the inputs, $w_i$ are the weights, $b$ is a bias term and $\sigma(\cdot)$ is the activation function. 
Lookup Table (LUT)-based DNNs (LUT-DNN) exploit FPGA-native elements to achieve exceptional area efficiency and latency. This is done by combining the multiplication, sum, and activation operations of Figure~\ref{fg:neuron} (Equation~\ref{eq:neuron} in a single LUT). Recent approaches, including LogicNets~\cite{LogicNets}, PolyLUT~\cite{polylut}, and NeuraLUT~\cite{neuralut}, introduce \textit{a priori} weight sparsity by constraining each neuron's fan-in, $F$. Put another way, they constrain the weight vector $W = \{{w_1},{w_2}, \ldots ,{w_N}\} \in {\mathbb{R}^{{N}}}\label{eq:W}$ to be determined \textit{a priori}, sparse and have a maximum of $F$ non-zero values ($0 < F \ll N$). 

The accuracy of existing LUT-based DNNs is fundamentally constrained by limited scalability and suboptimal random sparsity. To keep the truth table size tractable, both the input bit-width $\beta$ and the fan-in $F$ must remain small, significantly limiting the network’s representational capacity. In addition, the prevalent use of randomly assigned sparse connections often results in inefficient input selection, further impeding accuracy.

\begin{figure}[]
    \centerline{\includegraphics[width=0.99\linewidth]{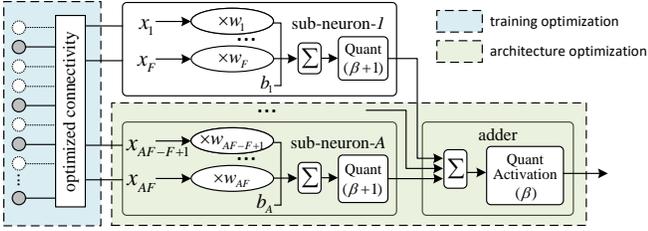}}
    \caption{Optimization strategies of SparseLUT} 
    \label{fg:sparselut}
\end{figure}

Although prior architectural enhancements have attempted to improve accuracy by boosting the expressive power of individual neurons, such as replacing linear operations with polynomial functions~\cite{polylut} or sub-network structures~\cite{neuralut}, these approaches still retain the fundamental limitation that the LUT size grows exponentially with $\beta F$, and thus fail to address the scalability bottleneck.
Meanwhile, existing sparsity optimization techniques, including post-training pruning~\cite{lecun1989optimal,thimm1995evaluating,gale2019state,sze2020efficient} and training-time sparsity methods~\cite{gmp,deepr,rigl,srigl}, are designed for conventional DNNs. They typically operate at the model or layer level and fail to address the unique neuron-level fan-in constraints specific to LUT-based networks. These limitations underscore a critical gap and an opportunity for innovation.

This paper introduces SparseLUT, an architecture that addresses the unique challenges outlined through architectural and training algorithmic level optimization. Figure~\ref{fg:sparselut} shows the separate optimization respectively. 

\begin{figure*}[t]
    \centerline{\includegraphics[width=0.90\linewidth]{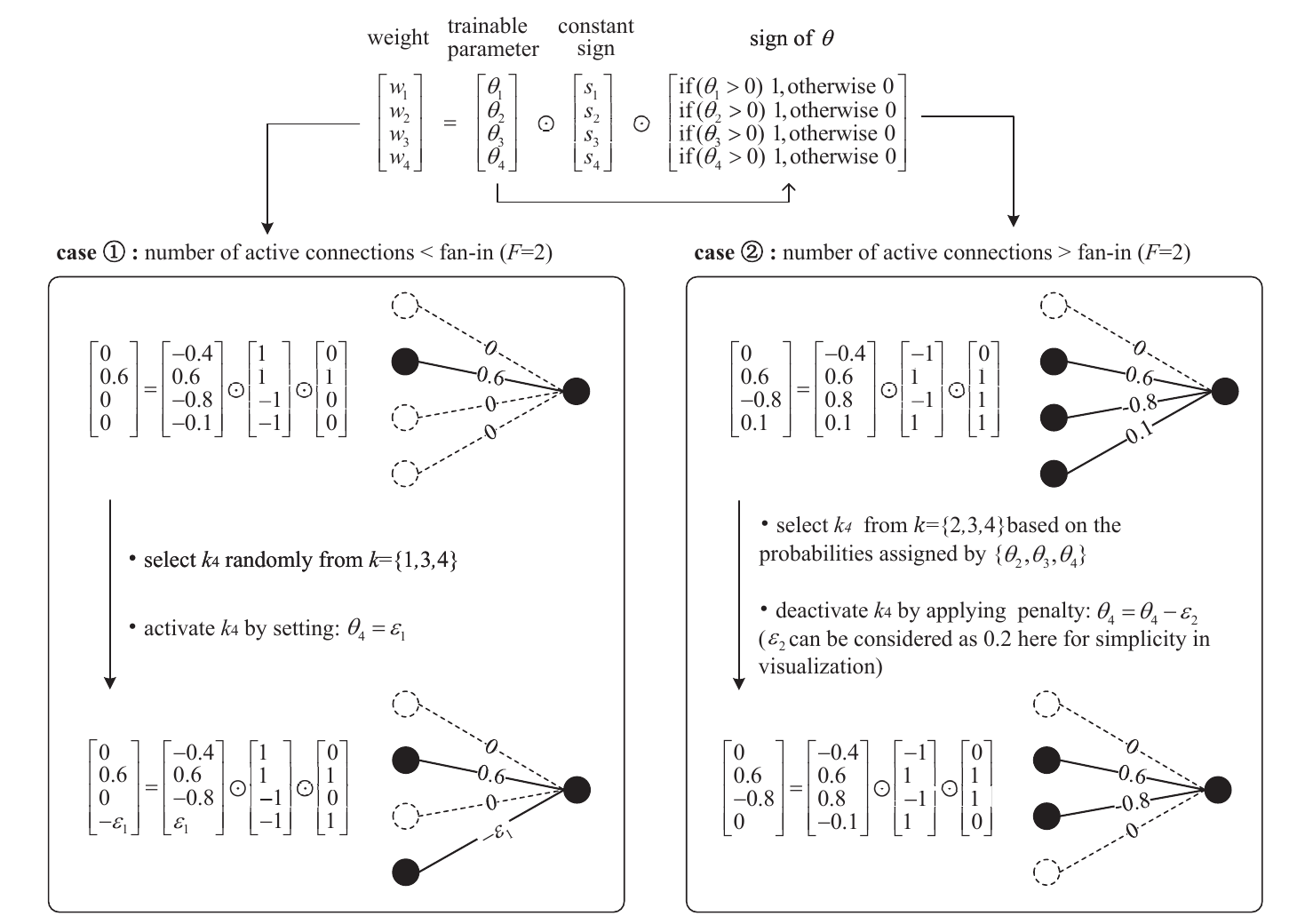}}
    \caption{Example iteration showing SparseLUT adjusting neuron connections to reach a target fan-in of 2, introducing new ones to increase the fan-in from 1 to 2 and eliminating connections to reduce the fan-in from 3 to 2. Note that the \( \epsilon_2 \) parameter is typically set to a much smaller value, requiring multiple iterations to deactivate a connection.} 
    \label{fg:example_f2}
\end{figure*}

First, at the architecture level, we combine $A$ copies of PolyLUT sub-neurons via an $A$-input adder to increase neuron fan-in. As shown in the right part of Figure~\ref{fg:sparselut}, our approach restructures the neuron computation: weight multiplication, accumulation, batch normalization (BN), and quantized activation to two stages. The first stage is similar to PolyLUT without the batch normalization and is repeated for each sub-neuron. Instead, the batch normalization is performed after the results are accumulated, with the resulting activation quantized again if necessary. In this way, we can make better use of the FPGA fabric, where the lookup table cost of $\mathcal{O}(2^{\beta FA})$ is reduced to $\mathcal{O}(A \times 2^{\beta F} + 2^{A(\beta +1)})$.

Second, we proposed a non-greedy sparsification strategy through careful selection of the non-zero $w_k$ values in Equation~\ref{eq:neuron}. Specifically, Figure~\ref{fg:example_f2}, to be elaborated in Section~\ref{se:Methodology}, illustrates our approach to training wherein active connections are gradually modified to ultimately achieve a target fan-in of $F=2$. As training progresses, new connections are added to neurons with fewer than \( F \) active connections (Case~\ding{172}), while neurons with active connections exceeding \( F \) (Case~\ding{173}) are deactivated. Over the training process, the target fan-in is achieved for all neurons.

The main contributions of this work are as follows: 

\begin{enumerate}

\item At the computer architecture level, we propose an efficient reconstruction and extension of the PolyLUT framework~\cite{polylut}, which integrates $A$ PolyLUT sub-neurons combined through an $A$-input adder, enhancing both scalability and accuracy.

\item At the tool level, we introduce the first non-greedy training algorithm for LUT-based neural networks that optimizes neuron input connectivity during training. This method imposes no additional LUT or routing resource overhead, as it solely modifies the selection strategy of the $F$ inputs depicted in Figure~\ref{fg:neuron}.

\item These two optimizations are orthogonal. In our evaluation, we first apply the proposed architectural optimization to the original PolyLUT. At a comparable accuracy level, our method reduces LUT consumption by 4.8$\times$ and 13.9$\times$ on the MNIST and Jet Substructure Classification tasks, respectively, while also decreasing latency by 1.2$\times$ to 1.6$\times$. Furthermore, our proposed connectivity optimization consistently enhances accuracy across baseline LUT-DNN models, including LogicNets~\cite{LogicNets}, PolyLUT~\cite{polylut}, and NeuraLUT~\cite{neuralut}, without incurring any hardware or latency overhead. It achieves up to a 2.13\% accuracy improvement on MNIST and a 0.94\% gain on Jet Substructure Classification compared to random sparsity.

\end{enumerate}

This work forms an extension of an earlier conference paper~\cite{polylutadd}; the key differences are the tool-level non-greedy training algorithm (Section~\ref{se:Sparsification Optimiztion}). The remainder of this paper is organized as follows: Section~\ref{se:Background} reviews related work, including LUT-based DNNs and deep sparsification techniques. Section~\ref{se:Methodology} describes the proposed method in detail, including the architecture-level optimization and non-greedy sparsification strategies. Section~\ref{se:Results} presents experimental results, showing accuracy and resource utilization improvements over existing methods. Finally, Section~\ref{se:Conclusion} concludes the paper and outlines future research directions.

\section{Background}
\label{se:Background}

\begin{figure*}
    \centerline{\includegraphics[width=0.80\linewidth]{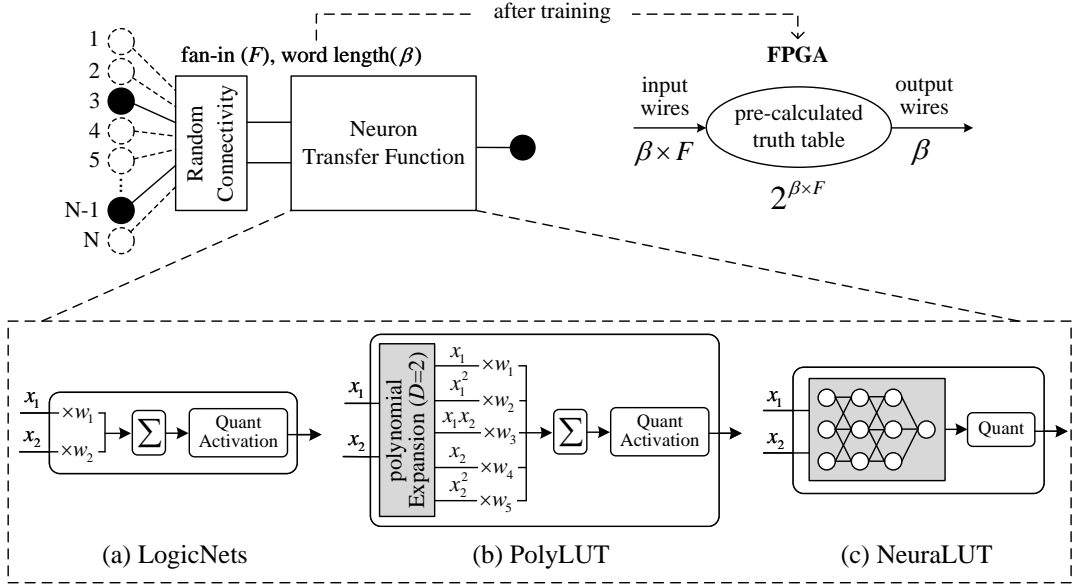}}
    \caption{Illustration of the generalized LUT-DNN architectures with sparse connectivity.} 
    \label{fg:lut}
\end{figure*}

Machine learning inference on FPGAs can be deployed through a variety of approaches. The most straightforward method is DSP-based inference, as exemplified by hls4ml~\cite{hls4ml}, which enables developers to train models in Python using popular frameworks such as TensorFlow or PyTorch and automatically generate synthesizable HLS code. This approach leverages the full range of FPGA resources, including DSP blocks. Another line of work is XNOR-based inference, such as Binary Neural Networks (BNNs)~\cite{bnn}, which employ 1-bit quantization to replace multipliers with efficient XNOR operations. Pure LUT-based methods can be further categorized into tree-based, differentiable LUT-based, and traditional DNN LUT-based approaches. For instance, TreeLUT~\cite{treelut} implements Gradient Boosted Decision Trees (GBDTs) using only LUTs without relying on BRAMs or DSPs, while Differentiable Weightless Neural Networks (DWNs)~\cite{dwn} construct neural networks entirely from weightless nodes mapped to LUTs. This work primarily targets traditional DNN-based LUT designs, while comparisons with other categories are provided in the evaluation section.

\subsection{LUT-based traditional DNN methods}

LogicNets~\cite{LogicNets} is the first LUT-based DNN architecture that quantizes the inputs and outputs of each neuron and encapsulates the neuron's transfer function ({\em i.e.}, densely connected linear and activation functions) in a lookup table. This method enumerated all possible combinations of a neuron's inputs and determined the corresponding outputs based on the neuron's weights and biases. By replacing popcount operations with Boolean expressions, significant computational savings were made. Building upon the foundations of LogicNets, PolyLUT~\cite{polylut}, proposed by Andronic {\em et al.}, further enhanced accuracy and reduced the number of required layers by introducing piecewise polynomial functions. Later, in the direction of continuing to improve the representation ability of each LUT, NeuraLUT~\cite{neuralut} employs the network-in-network technique~\cite{nin} and, therefore, maps entire sub-nets to a single LUT, enabling a deeper NN and better accuracy.

Figure~\ref{fg:lut} illustrates the main idea behind these LUT-DNN approaches. As shown in the top part, a maximum of $F$ inputs are randomly selected from ${N}$ nodes of the current layer to connect to each neuron of the next layer (only a single neuron of the next layer is demonstrated here). Additionally, the bit width of each neuron's inputs and outputs is quantized as $\beta$, and the rest of the parameters inside the neuron transfer function are maintained in full precision. Therefore, after training, the transfer function mapping an input vector $\left[ {{x_1},{x_2}, \ldots ,{x_{F}}} \right]$ to the output node can be implemented using $\beta F$ inputs in hardware, and hence its implementation requires $\mathcal{O}(2^{\beta F})$ LUTs in a pre-calculated truth table.

\subsection{Sparsification}

\begin{table*}[t]
\centering
  \caption{Summary of dynamic sparsification approaches.}
  \def\arraystretch{1.25}
  \setlength{\tabcolsep}{0.3cm}
  \resizebox{0.98\linewidth}{!}{ 
  \begin{tabular}{|c|c|c|c|c|}
  \hline
  \multicolumn{1}{|c|}{{Method}}    & \multicolumn{1}{c|}{{Training}} & \multicolumn{1}{c|}{{Drop (Prune) $\searrow$}} & \multicolumn{1}{c|}{{Regrowth $\nearrow$}} & \multicolumn{1}{c|}{{\textit{a priori} fixed fan-in constrain}}\\ \hline \hline
  GMP~\cite{gmp}       & dense-to-sparse   & magnitudes  & \ding{55}    & \ding{55}   \\
  \hline
  DeepR~\cite{deepr}   & sparse-to-sparse  & stochastic  & random      & \ding{55}   \\
  \hline
  RigL~\cite{rigl}     & sparse-to-sparse  & magnitudes  & gradient    & \ding{55}   \\
  \hline
  SRigL~\cite{srigl}   & sparse-to-sparse  & magnitudes  & gradient    & \ding{55}   \\
  \hline
  DWN~\cite{dwn}       & sparse-to-sparse  & magnitudes  & \ding{55}    & \ding{51}$^*$   \\
  \hline
  SparseLUT (ours)    
  & flexible          & stochastic+magnitudes  & random  & \ding{51}   \\
  \hline
\end{tabular}}
\label{tb:dst_review}

\begin{tablenotes}
    \footnotesize
    \item[1] $^*$ DWN is a special case here, designed specifically for weightless neural networks, and is not applicable to generalized weighted DNNs.
\end{tablenotes}

\end{table*}

In practice, for most dense DNNs, a significant fraction of weights are redundant and can be set to zero with minimal effect on accuracy~\cite{sze2020efficient}. Research on sparsifying neural networks, known as network pruning, dates back to the late 1980s and mid-1990s. These techniques begin with a pre-trained, densely connected model and perform weight removal, where unimportant weights are identified and set to zero. Various approaches have been used as the removal criteria. For example, early work in 1989, Optimal Brain Damage~\cite{lecun1989optimal}, removed non-salient weights as determined via the diagonals of the Hessian matrix. In 1995, Thimm \emph{et al.}~\cite{thimm1995evaluating} demonstrated that removing weights based on their magnitude was a simple yet effective technique. Gale \emph{et al.}~\cite{gale2019state} further examined alternative pruning criteria on large-scale learning of transformers trained on WMT2014 English-to-German and ResNet-50 trained on ImageNet. They concluded that complex techniques such as $L_0$ regularization~\cite{louizos2017learning} and Variational Dropout~\cite{molchanov2017variational} achieve similar accuracy-sparsity trade-offs compared to magnitude-based pruning. As for the LUT-DNN scope, PolyLUT~\cite{polylut2} and NeuraLUT-Assemble~\cite{neuralut-assemble} introduce a post-training pruning strategy based on weight ranking to replace the random connectivity. Specifically, after training, for each neuron, the weight branches are ranked according to their fan-in, and the top branches are selected in descending order of importance.

Gale \emph{et al.}~\cite{gale2019state} also observed that unstructured sparse architectures learned through pruning alone cannot achieve as good performance as a model trained with joint sparsification and optimization. Thus, more recent methods commence from scratch rather than a pre-trained model, \emph{i.e.}, they (1) start from the original initial training conditions and (2) update connectivity by doing weight removal and weight re-growth (optional) simultaneously.
This can be further divided into two subcategories with representatives listed in Table~\ref{tb:dst_review}: (1) The training starts with a dense model and gradually removes connections during training to achieve the desired sparsity~\cite{gmp}. (2) The training begins with a sparse model to minimize training memory overhead, and during training, connections are dropped and regrown, guided by criteria such as weight magnitudes, gradients, or stochastically~\cite{rigl,srigl,deepr}.

Gradual Magnitude Pruning (GMP)~\cite{gmp}, removes connectivity from a dense initial solution during training until the desired sparsity is reached. It begins with a dense network and removes weights with the smallest magnitudes. Its sparsity decreases monotonically until the target sparsity is reached. 

In contrast, there are approaches that employ dynamic rewiring of neural networks~\cite{deepr, rigl, srigl}. This strategy is inspired by the human brain, where the majority of brain volume is occupied by white matter—the network of connections between neurons. In the brain, synaptic connectivity is highly dynamic, with the continuous formation and reorganization of synapses, particularly during learning~\cite{holtmaat2005transient, stettler2006axons, attardo2015impermanence, chambers2017stable}.

Deep Rewiring (DeepR)~\cite{deepr} introduces a stochastic framework for dynamic rewiring. At initialization, DeepR employs a network with a limited number of random connections. During training, at each iteration, connections are removed if their weight updates lead to a sign change. A new connection is then randomly activated to preserve the target sparsity level. Our work extends this random walk mechanism, and further details are provided in subsequent sections.
Rigged Lottery (RigL)~\cite{rigl} removes weights with the smallest magnitudes and regrows connections based on gradient magnitude during training. At each iteration, the number of dropped connections equals the number of regrown connections, maintaining a consistent sparsity level. RigL is designed to handle standard unstructured sparsity constraints at the layer level.
Structured RigL (SRigL)~\cite{srigl} extends RigL by incorporating constant fan-in structured sparsity, which facilitates a compact weight matrix representation. Specifically, after training, the non-zero weights in each matrix exhibit a rectangular pattern. SRigL achieves its target fan-in target on average, rather than this being enforced on a per-neuron basis. 

DWN's element-wise weight-free mapping is not directly applicable to our LUT-based weighted inference pipeline
DWN~\cite{dwn} is a weightless neural network model, where its learnable mapping mechanism selects a fixed number of connections based on the largest weights during training. Specifically, its computation does not involve weighted summation, accumulation, nonlinear activation, or standard backpropagation. Instead, it provides a differentiable mechanism to decide the routing of inputs directly into LUTs, enabling training without conventional weights.
Therefore, although DWN provides an insightful online learnable mapping, its weight-free, element-wise formulation is fundamentally different from the weighted computations underlying traditional LUT-based DNN inference, particularly for networks with customized architectures and training pipelines (e.g., the polynomial formulation in PolyLUT and the network-in-network structure in NeuraLUT). As a result, its direct integration into a conventional weighted-DNN training workflow has not yet been demonstrated in the existing literature.

\section{Methodology}
\label{se:Methodology}

\subsection{DNN Architecture Optimization}
\begin{figure}
    \centerline{\includegraphics[width=1.0\linewidth]{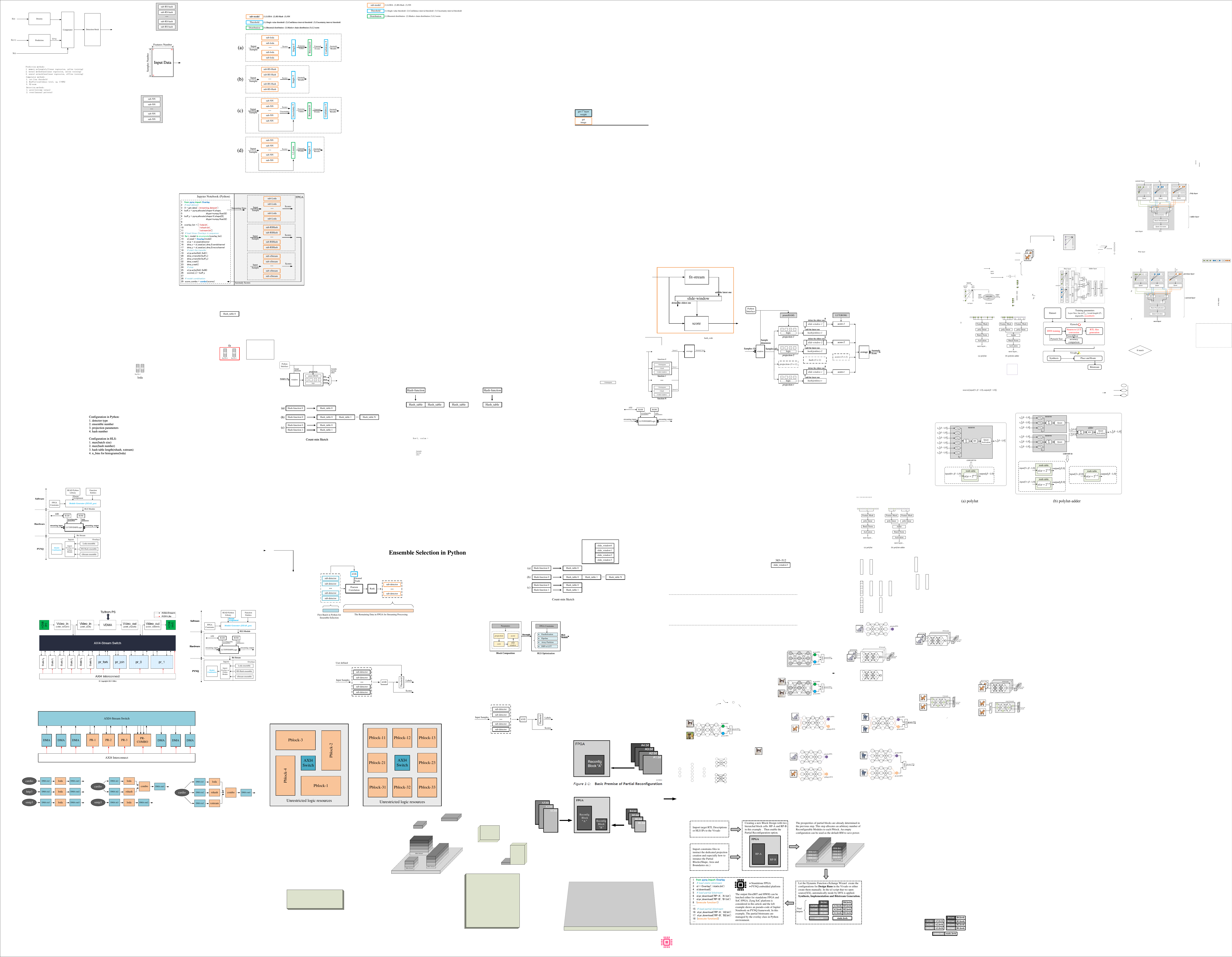}}
    \caption{A single-layer block diagram of the optimized architecture.}
    \label{fg:layer_structure}
\end{figure}

Figure~\ref{fg:layer_structure} outlines our proposed DNN architecture. Compared with Figure~\ref{fg:lut}, the fan-in $F$ to sub-neurons remains the same, but the total fan-in to the neuron is increased by a factor of $A$ at the output. This is achieved by summing $A$ independent and parallel randomly connected Poly-layers. The connectivity between the Poly-layer and the Adder-layer is fixed to perform the addition operation for each neuron.
To elucidate the enhancement mechanism, we introduce the formulation detailed in Eq.~\eqref{eq:equal}.

\begin{small}
\begin{equation}
    \label{eq:equal}
    \sum\limits_{i = 0}^{AF - 1} {{w_i}{x_i} + b}  = \sum\limits_{a = 0}^{A - 1} {\left( {\sum\limits_{i = 0}^{F - 1} {{w_{(aF + i)}}{x_{(aF + i)}}}  + {b_a}} \right)} 
\end{equation}
\end{small}

During computation, the activation function, such as Rectified linear unit (ReLU) outputs bits, can be one bit less than the input bits because its output is non-negative. To avoid overflow in the Adder-layer, we increase its internal word length by one bit (to $\beta+1$), as seen in Figure~\ref{fg:sparselut}.

\subsection{Dynamic Sparsification}
\label{se:Sparsification Optimiztion}
\subsubsection{Model Representation}

Unlike conventional DNNs, which focus solely on updating the network weights and biases, SparseLUT also incorporates weight removal and regrowth under a set of predefined constraints. A key factor enabling this is its weight representation. We follow the approach of DeepR~\cite{deepr} where each connection \( k \) in the network is represented by two components:

\begin{itemize}
    \item A \textbf{trainable connection parameter} $\theta_k \in \mathbb{R}$, which controls both the magnitude of the connection and its status (active or inactive).
    \item A \textbf{predefined, non-trainable constant sign} $s_k \in \{-1, 1\}$, initialized randomly, which encodes the fixed polarity of the connection.
\end{itemize}

Referring to Figure~\ref{fg:example_f2}, we use $W = \{{w_1},{w_2}, \ldots ,{w_N}\} \in {\mathbb{R}^{{N}}}$ to denote the weight vector of a neuron in Equations~\ref{eq:neuron}.

Algorithm~\ref{alg:sparselutweight} outlines the procedure for determining the effective weight vector for each neuron in SparseLUT. The algorithm begins with an initial weight vector \( W_0 \in \mathbb{R}^N \), where the magnitude of each element is used as the initial value for \(\theta_k\). This ensures dense connectivity under the initial conditions, with all connections corresponding to \(\theta_k\) being non-negative. The initial sparsity of the network is determined by the element-wise product of \(|W_0|\) and a binary sparsity mask vector \(\textit{is\_con}\), where only the element $F_i$ is set to 1, and all other elements are set to 0. To finalize the effective weight vector \( W \), an indicator function \(\text{1}(\cdot)\) is applied to the product \(\theta_k \odot s_k\), controlling the active status of each connection.

\begin{algorithm}
\caption{Weight Mapping Procedure}
\label{alg:sparselutweight}
\KwIn{Weight dimensions for $W$, initial fan-in $F_i$.}
\BlankLine
$W_0 \gets$ Randomly initialize weights $W_0$ from a standard normal distribution with the same dimensions as $W$\;
$is\_con \gets$ Generate a binary sparsity mask $is\_con$ such that each output neuron is connected to $F_i$ randomly chosen input neurons\;
$\theta \gets |W_0| \odot is\_con$ {(Compute the initial $\theta$ values as the element-wise product)}\;
$sign \in \{-1, +1\}$ Randomly initialize weight signs with uniform probability\;
$W \gets \theta \cdot sign \cdot \text{1}(\theta > 0)$ {(Compute the final weights, where $\text{1}(\cdot)$ is the indicator function that sets negative values of $\theta$ to zero)}\;
\BlankLine
\KwRet{Final weight vector $W$}
\end{algorithm}

\subsubsection{Sparse Training}

\begin{algorithm}
\caption{SparseLUT Training Procedure}
\label{alg:sparselut}
\KwIn{Target fan-in $F_o$, sparse-to-sparse training starting point $T$, learning rate $\eta$, regularization coefficient $\alpha$, and noise ${v_k} \sim N(0,G^2)$.}
\KwOut{Feature mask $\mathcal{M}$}
\BlankLine
\textbf{Initialization:}\\
$W \gets $ defined in Algorothm~\ref{alg:sparselutweight}\;
\BlankLine
\textbf{Training:}\\
\For{each training step $t$}{
    
    \For{all active connections $k$ ($\theta_k>0$)}{
        Update $\theta_k \leftarrow \theta_k - \eta \frac{\partial E(\theta)}{\partial \theta_k} - \eta \alpha + \eta {v_k}$\;
        \textbf{if} $\theta_k < 0$ \textbf{then} set connection $k$ non-active \;
    }
    $R \leftarrow$ number of active connections - $F_o$\;
    \eIf{$R<0$}{
        Select $\left| R \right|$ non-active connections $k'$ with uniform probability and activate them\;
        Set $\theta_{k'} \leftarrow \epsilon_{1}$\;
    }
    {
        \eIf{$t<T$}{
            Rank the \( \theta \) values of all active connections\;  
            Select \( \left| R \right| \) active connections \( k' \) with the lowest rank probabilities and apply penalties\;
            Update ${\theta _{k'}} \leftarrow {\theta _{k'}} - \epsilon_{2}$\;
        }
        {
            Rank the \( \theta \) values of all active connections\;  
            Select $\left| R \right|$ active connections $k'$ with the lowest rank probabilities and deactivate them\;
            Set ${\theta _{k'}} \leftarrow 0$\;
        }
    }
}
\BlankLine
\KwRet Feature mask $\mathcal{M}$, where $\mathcal{M}_k = 1$ if $\theta_k > 0$, otherwise $\mathcal{M}_k = 0$\;
\end{algorithm}

The SparseLUT training procedure, as outlined in Algorithm~\ref{alg:sparselut}, is a non-greedy algorithm. Each iteration updates the connection parameters and adjusts the network's sparsity to enforce the target fan-in \( F_o \) sparsity during training.

At each backpropagation training step~\cite{Sun_2019_sgdsurvey}, only the parameters \( \theta_k \) of active connections (\( \theta_k > 0 \)) are updated according to the rule in Line~6. Given that the learning rate is $\eta$, the term \( \eta \frac{\partial E(\theta)}{\partial \theta_k} \) represents the derivative of the error function, while \( \eta \alpha \) is regularization term ($\alpha$ is any regularization function). The final term, \( \eta {v_k} \), originally introduced in DeepR~\cite{deepr}, implements a random walk scheme in the parameter space. Here, \( {v_k} \sim N(0, G^2) \) denotes a noise matrix with the same dimensions as \( \theta_k \), a mean of \( 0 \), and a standard deviation of \( G \). Following this update, any connection for which \( \theta_k \) becomes negative is deactivated.

For each neuron, the difference between the existing number of active connections and the target fan-in is computed as \( R \). A negative \( R \) indicates that the number of active connections is smaller than the target fan-in, prompting the activation of \( |R| \) inactive connections \( k' \) selected uniformly at random (Line~10). These connections are initialized with a small value \( \epsilon_1 \) to allow them to rejoin the backpropagation process without dominating the updates immediately after reactivation.

Conversely, a positive \( R \) means that \( |R| \) active connections need to be deactivated to meet the fan-in constraint. Different from approaches in Table~\ref{tb:dst_review}, SparseLUT employs a two-phase training strategy for this deactivation process:  

\begin{enumerate}
    \item {progressive sparsification phase:}  
    During the early stages of training (\( t < T \)), a small penalty term \( \epsilon_2 \) is subtracted from \( \theta_{k'} \) in each iteration (Line~16). This gradual penalty reduces the values of smaller parameters to a greater extent, under the assumption that connections with smaller weights contribute less to the neuron’s output. The speed of connectivity convergence is significantly influenced by the \( \epsilon_2 \). As \( \theta_{k'} \) becomes negative, these connections are eliminated. Specifically, a large value of $\epsilon_{2}$ can cause nodes with relatively larger weights to change state immediately. This stage relaxes the strict requirement that the number of dropped and regrown connections must always match, as used in conventional regrowth approaches such as DeepR~\cite{deepr}, RigL~\cite{rigl}, and SRigL~\cite{srigl}. We will later show that this greater flexibility during training leads to consistently improved accuracy.

    \item {fine-tuning phase:}  
    In the later stage of training (\( t \geq T \)), the model strictly enforces the fan-in constraint. Connections are directly deactivated when \( |R| > 0 \) (Line~20). This phase focuses on fine-tuning the remaining connections through periodically rewiring eliminated connections and selecting the same number of alternatives to maintain the fan-in constraint for each neuron.  
\end{enumerate}

Upon completing the training process, the final generated output of the algorithm is formed as the feature mask \( \mathcal{M} \), where each element \( \mathcal{M}_k \) indicates the status of connection \( k \) (Line~21). 

The computational cost of SparseLUT, as shown in Algorithm~\ref{alg:sparselut}, primarily arises from the connectivity search, specifically the Prune and Regrowth steps. When $R<0$ (regrowth), selection is performed from all inactive candidates ($N - F \approx N$), while for $R>0$ (pruning), ranking is conducted on active candidates. Empirically, the number of active candidates quickly stabilizes near the target fan-in value $F_o$ during extended training periods. As a result, the ranking overhead mainly depends on $F_o$. Therefore, the training time is primarily influenced by the model size, the number of epochs, and $F_o$.

\subsection{ToolFlow}

\begin{figure}[h]
    \centerline{\includegraphics[width=0.66\linewidth]{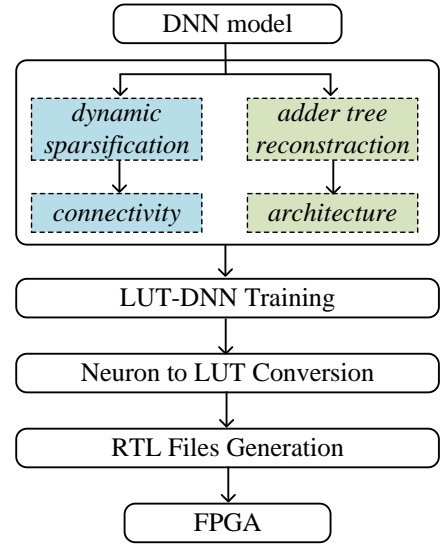}}
    \caption{Workflow of SparseLUT.} 
    \label{fg:toolflow}
\end{figure}

Figure~\ref{fg:toolflow} illustrates the overall workflow of the proposed SparseLUT framework, which is readily applicable to existing LUT-based DNN designs~\cite{LogicNets,polylut,neuralut}. The process begins with the same model configuration as standard LUT-DNNs, thereby maintaining full compatibility with their training pipelines. On the left side of the flow, SparseLUT generates a \emph{connectivity feature mask} \( \mathcal{M} \), which conforms to the structural requirements of existing LUT-DNNs and directly replaces the randomly generated connectivity shown in Figure~\ref{fg:lut}. In addition, the network architecture is modified to incorporate the fan-in factor \( A \) as a configurable parameter. This enables the reconstruction of the network topology using an adder tree structure.

Following this, LUT-DNN training is conducted offline using PyTorch~\cite{PyTorch}. Integration with the Brevitas library~\cite{brevitas} enables quantization-aware training (QAT), ensuring that the trained model is suitable for hardware deployment. Post-training, the model's weights are transformed into lookup tables. This transformation begins by analyzing the quantized activations to determine each neuron's input range. For sub-neuron layers, all possible input combinations are generated based on the parameters \( \beta \) and \( F \); for adder layers, combinations are generated using \( \beta \) and \( A \). These input vectors are then propagated through their respective layers to obtain output values, forming input--output pairs that constitute the contents of the lookup tables.

These lookup tables are subsequently used to generate Register Transfer Level (RTL) code in Verilog, capturing the Boolean functions implemented by each neuron. Finally, the design is synthesized onto hardware using the AMD/Xilinx Vivado toolchain~\cite{Vivado}.

\section{Results}
\label{se:Results}

\subsection{Datasets}
We evaluated the proposed SparseLUT design using two widely adopted datasets for ultra-low latency inference. These datasets align with the evaluation benchmarks of existing techniques, including LogicNets~\cite{LogicNets}, PolyLUT~\cite{polylut}, and NeuraLUT~\cite{neuralut}. In addition, to evaluate the generalizability of the optimized connectivity on more complex tasks, we also include an ablation study on the classification accuracy with the CIFAR-10 dataset. Since all the above LUT-DNN methods didn't conduct any CIFAR-10-based experiment and our focus is not on exploring the optimal hyperparameter search space here, the setups used for MNIST are directly applied to the CIFAR-10 task.

\begin{enumerate}
\item \textit{Handwritten Digit Recognition}: In timing-critical sectors such as healthcare, medical imaging, and real-time object tracking, low latency is crucial. These applications underscore the necessity for swift and accurate decision-making, where even minimal delays can have significant repercussions. Unfortunately, there is no public dataset specialized for low-latency image classification tasks, so the MNIST~\cite{mnist} is utilized to compare our work with other LUT-based DNNs. MNIST is a dataset for handwritten digit recognition tasks with $28 \times 28$ pixels as input and 10 classes as outputs.

\item \textit{Jet Substructure Classification}:  Real-time decision-making is often important for physics experiments such as the CERN Large Hadron Collider (LHC). Jet Substructure Classification (JSC) is one of its applications that requires high-throughput data processing. Prior works~\cite{ngadiuba2020compressing,duarte2018fast,coelho2021automatic,fahim2021hls4ml} employed neural networks on an FPGA for this task to provide real-time inference capabilities. We also use the JSC dataset formulated from Ref.~\cite{duarte2018fast} to evaluate our work, with the dataset having 16 substructure properties as input and 5 types of jets as outputs. FPGA-based classification in this task must be pipelined to manage a data rate of 40 MHz while ensuring the response latency remains under a microsecond.

\item \textit{Object Recognition}: Similar to handwritten digit recognition, object recognition in real-world scenarios also demands low latency to enable fast and reliable decision-making. To evaluate our method, the CIFAR-10~\cite{cifar10} dataset is employed, which provides a more complex benchmark compared to MNIST. CIFAR-10 consists of $32 \times 32$ color images from 10 diverse object categories, making the classification task more challenging.  

\end{enumerate}

We emphasize that the architectural optimization and the connectivity training optimization proposed in SparseLUT are orthogonal and can be applied independently. To facilitate an ablation study, we first evaluate the model that incorporates only the architectural optimization. For consistency with the terminology used in our published work~\cite{polylutadd}, we refer to this architecture-only variant as PolyLUT-Add.

\subsection{Architectural Optimization Evaluation}

Using AdamW as the optimizer~\cite{PyTorch}, we trained the smaller models/datasets (JSC-M Lite) for 1000 epochs and used 500 epochs for (JSC-XL and HDR). The mini-batch size is set to 1024 and 128 for (JSC-XL, JSC-M Lite) and MNIST, respectively. 
We inherit these configurations from PolyLUT~\cite{polylut} to ensure consistency in evaluation. 

The hardware evaluation is compiled using Vivado 2020.1 on the \texttt{xcvu9p-flgb2104-2-i} FPGA part; the \texttt{Flow\_PerfOptimized\_high} settings were configured to perform synthesis in the \texttt{Out-of-Context} (\texttt{OOC}) mode, and Default values were used as the Place \& Route settings.

\subsubsection{PolyLUT-Add vs. original PolyLUT}

\begin{table*}[]
  \centering
  \caption{Comparison of accuracy and hardware results between PolyLUT and PolyLUT-Add ($\mathbb{D}=1$, $\mathbb{W}=1$)}
  \label{tb:resuts_3datasets}
  \bgroup
  \def\arraystretch{1.2}
  \setlength\tabcolsep{1mm}
  \scalebox{0.99}{
  \begin{tabular}{|c|c|l|c|c|l|cccc|c|}
  \cline{1-11}
  \multirow{2}{*}{{Models}}  & \multirow{1}{*}{{Degree}} & \multicolumn{1}{c|}{\multirow{2}{*}{{Model}}} & {Fan-in} & \multirow{2}{*}{{Acc(\%)$\uparrow$}} & \multicolumn{1}{c|}{ {lookup table}} & {LUT}          & {FF} & {$F\_max$} & {Latency}  & {RTL Gen.}                 \\ 
   & $D$ & & {$ (F \times A)$} & & \multicolumn{1}{c|}{ {entries$\downarrow$} }    &  {(\% of 1182240)}          & {(\% of 2364480)} & {(MHz)} & {(cycles)}   & {(hours)} \\ \hline\hline
   \multirow{8}{*}{HDR}  & \multirow{4}{*}{1} & \multirow{2}{*}{PolyLUT}        & 6          & 93.8       & $2^{12}$                     & 3.43& 0.12 & 378 & 6 & 1.40  \\
                           &                    &                                  & 10         & 96.1      & $2^{12} \times 256$          & \multicolumn{4}{c|}{ \(-\) } & \(-\)  \\ \cline{3-3}
                           &                    & \multirow{2}{*}{PolyLUT-Add}     & 6$\times$2 & 96.5       & $2^{12} \times 2$ + $2^{6}$  & 12.69 & 0.12 & 378 & 6 & 3.00  \\
                           &                    &                                  & 6$\times$3 & \bf{96.6}  & $2^{12} \times 3$ + $2^{9}$  & 20.67 & 0.12 & 378 & 6 & 4.40  \\ \cline{2-11}
                           & \multirow{4}{*}{2} & \multirow{2}{*}{PolyLUT}         & 6          & 95.4       & $2^{12}$                     & 6.62 & 0.12 & 378 & 6 & 1.40  \\
                           &                    &                                  & 10         & 97.3       & $2^{12} \times 256$           & \multicolumn{4}{c|}{\(-\)} & \(-\) \\ \cline{3-3}
                           &                    & \multirow{2}{*}{PolyLUT-Add}     & 6$\times$2 & 97.1       & $2^{12} \times 2$ + $2^{6}$  & 19.78 & 0.07 & 378 & 6 & 3.00  \\
                           &                    &                                  & 6$\times$3 & \bf{97.6}  & $2^{12} \times 3$ + $2^{9}$  & 31.36 & 0.07 & 378 & 6 & 4.50  \\ \hline\hline
  
   \multirow{6}{*}{JSC-XL}  & \multirow{3}{*}{1} & \multirow{2}{*}{PolyLUT}       & 3          & 74.5       & $2^{15}$                  & 19.55& 0.07& 235 & 5 & 2.10  \\
                           &                    &                                  & 5          & 74.9       & $2^{15} \times 1024$        & \multicolumn{4}{c|}{ \(-\) } & \(-\)  \\ \cline{3-3}
                           &                    & \multirow{1}{*}{PolyLUT-Add}     & 3$\times$2 & \bf{75.1}  & $2^{15} \times 2$ + $2^{12}$ & 50.10 & 0.07 & 235 & 5 & 5.17  \\ \cline{2-11}
                           & \multirow{3}{*}{2} & \multirow{2}{*}{PolyLUT}         & 3          & 74.9       & $2^{15}$                     & 37.40 & 0.07 & 235 & 5 & 2.30  \\
                           &                    &                                  & 5          & 75.2       & $2^{15} \times 1024$         & \multicolumn{4}{c|}{ \(-\) } & \(-\)  \\ \cline{3-3}
                           &                    & \multirow{1}{*}{PolyLUT-Add}     & 3$\times$2 & \bf{75.3}  & $2^{15} \times 2$ + $2^{12}$ & 89.60 & 0.07 & 235 & 5 & 5.24  \\ \hline\hline

   \multirow{8}{*}{JSC-M Lite}  & \multirow{4}{*}{1} & \multirow{2}{*}{PolyLUT}   & 4          & 71.6       & $2^{12}$                     & 0.97 & 0.01 & 646 & 3 & 0.16  \\
                           &                    &                                  & 7          & 72.1       & $2^{12} \times 512$            & \multicolumn{4}{c|}{ \(-\) } & \(-\)  \\ \cline{3-3}
                           &                    & \multirow{2}{*}{PolyLUT-Add}     & 4$\times$2 & 72.2       & $2^{12} \times 2$ + $2^{8}$  & 2.62 & 0.01 & 488 & 3 & 0.35  \\
                           &                    &                                  & 4$\times$3 & \bf{72.3}  & $2^{12} \times 3$ + $2^{12}$ & 4.33& 0.01& 363 & 3 & 0.63  \\ \cline{2-11}
                           & \multirow{4}{*}{2} & \multirow{2}{*}{PolyLUT}         & 4          & 72.0       & $2^{12}$                     & 1.51& 0.01& 568 & 3 & 0.16  \\
                           &                    &                                  & 6          & 72.5       & $2^{12} \times 512$          & \multicolumn{4}{c|}{ \(-\) } & \(-\)  \\ \cline{3-3}
                           &                    & \multirow{2}{*}{PolyLUT-Add}     & 4$\times$2 & 72.5       & $2^{12} \times 2$ + $2^{8}$  & 4.29& 0.01& 440 & 3 & 0.34  \\
                           &                    &                                  & 4$\times$3 & \bf{72.6}  & $2^{12} \times 3$ + $2^{12}$ & 6.57& 0.01& 373 & 3 & 0.64  \\ \hline
  \end{tabular}}
  \egroup

  \begin{tablenotes}
    \footnotesize
    \item[1] \(-\): Data for very high fan-in settings is omitted due to exceeding FPGA memory capacity limits.
\end{tablenotes}
\end{table*}

As shown in Table~\ref{tb:resuts_3datasets}, for $A=2$, PolyLUT-Add achieves accuracy improvements of 2.7\%, 0.6\%, and 2.3\% over PolyLUT on the MNIST and Jet Substructure classification benchmarks, respectively, with only a 2–3$\times$ increase in LUT size. In contrast, achieving similar accuracy gains in PolyLUT by increasing the fan-in $F$ results in a 256–1024$\times$ increase in LUT size. These results demonstrate that PolyLUT-Add offers a more efficient trade-off between accuracy and hardware cost. Furthermore, it's noteworthy that the RTL Generation time cost also correlates with the number of lookup table entries; it follows that a direct increase in fan-in would incur exponentially higher RTL Generation time costs. (The RTL Generation time was measured on a desktop with Intel(R) Core(TM) i7-10700F @2.9GHz and 64GB memory.)

\subsubsection{PolyLUT-Add vs. Deeper and Wider PolyLUT}
\label{se:Accuracy}

The accuracy comparison between PolyLUT-Add and the other two variants of PolyLUT is further tested (the baseline PolyLUT setup is shown in Table~\ref{tb:polylutsetup}):

\begin{enumerate}
\item \textbf{PolyLUT-Deeper}: This explores the impact of increasing network depth. We denote the depth factor as $\mathbb{D}$. Then $\mathbb{D}\times$ the number of layers is applied to models in Table~\ref{tb:polylutsetup}. For example, for JSC-M Lite, if $\mathbb{D}=2$, the hidden layer is doubled, meaning the neurons per layer become (64,64,32,32,5). 

\item \textbf{PolyLUT-Wider}: This examines the impact of a wider network model. We denote the width factor as $\mathbb{W}$. Then $\mathbb{W}\times$ the number of neurons per layer is applied to models in Table~\ref{tb:polylutsetup}. Once again, for JSC-M Lite, if $\mathbb{W}=2$, the neurons per layer become (128,64,5).
\end{enumerate}

\begin{table}[]
  \centering
  \caption{Setups of PolyLUT baseline models for Figure~\ref{fg:acc_results}.}
  \label{tb:polylutsetup}
  \bgroup
  \def\arraystretch{1.16}
  \setlength{\tabcolsep}{0.3cm}
  \resizebox{0.99\linewidth}{!}{ 
  \begin{tabular}{|c|c|l|c|l|}
  \hline
{Model} & {Neurons per layer} & \textbf{$\beta$} & \textbf{$F$} & \textbf{$D$} \\ \hline \hline
  \multirow{1}{*}{MNIST}       & 256, 100, 100, 100, 100, 10 & 2       &  6  & 1,2 \\ \hline
  \multirow{1}{*}{JSC-XL$^1$}  & 128, 64, 64, 64, 5          & 5       &  3  & 1,2  \\ \hline
  \multirow{1}{*}{JSC-M Lite}  & 64, 32, 5                   & 3       &  4  & 1,2 \\  \hline

  \end{tabular}}
  \egroup

  \begin{tablenotes}
    \footnotesize
    \item[1] PolyLUT with $D=1$ is equivenlent to LogicNets.
    \item[2] Remarks:  $^1$ $\beta_{i}=7$, $F_{i}$ = 2;  
\end{tablenotes}
\end{table}

\begin{figure}[t]
    \centering
    \subfigure[HDR]{
    \hspace{-0.05\linewidth}
        \centering
        \label{fg:mnist}
        \includegraphics[width=0.89\linewidth]{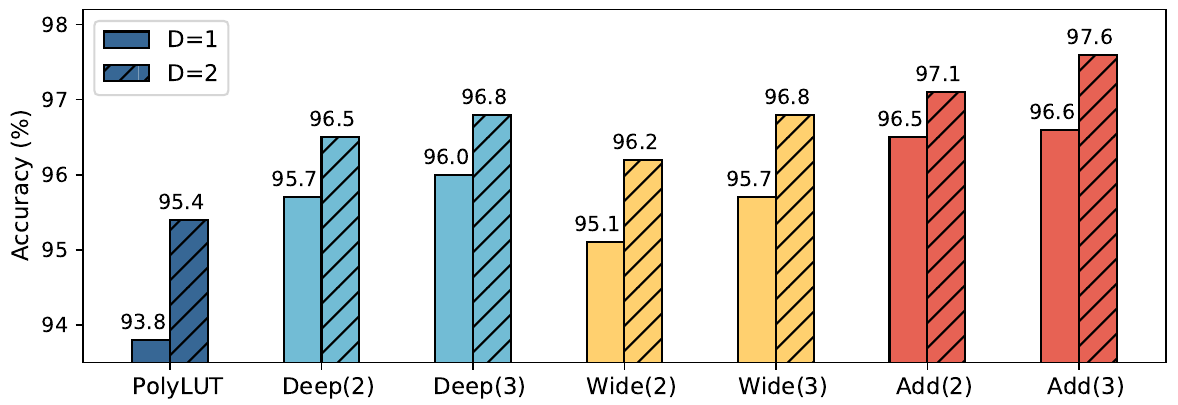}
    }\vspace{-2.3mm}
    \subfigure[JSC-XL]{
    \hspace{-0.05\linewidth}
        \label{fg:jsc-xl}
        \includegraphics[width=0.5\linewidth]{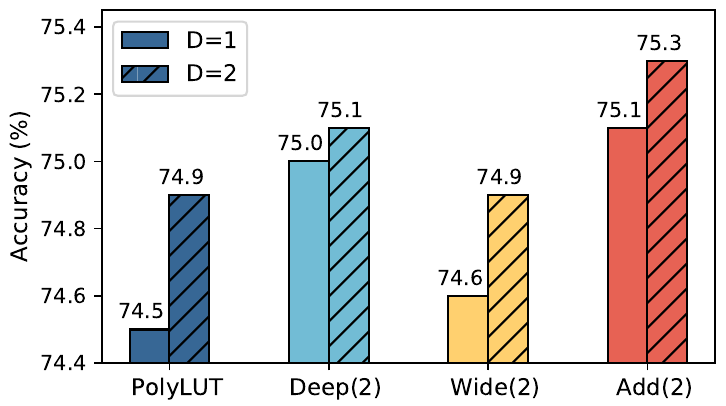}
    }\vspace{-2.3mm}
    
    \subfigure[JSC-M Lite]{
    \hspace{-0.05\linewidth}
        \centering
        \label{fg:jsc-mlite}
        \includegraphics[width=0.89\linewidth]{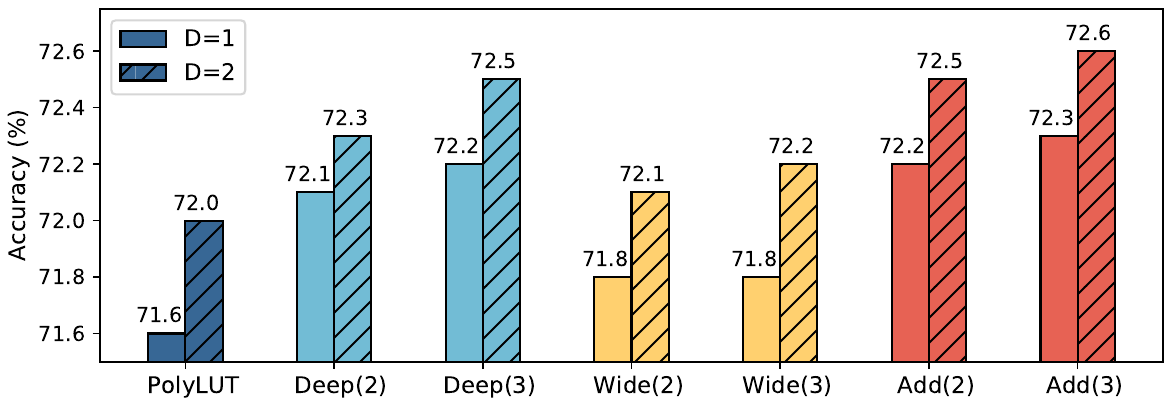}
    }\vspace{-2.3mm}
    \caption{Accuracy results on different models. We use Deep($\mathbb{D}$), Wide($\mathbb{W}$) and Add($A$) to denote ``PolyLUT-Deeper'', ``PolyLUT-Wider'' and ``PolyLUT-Add'' respectively.}
    \label{fg:acc_results}
\end{figure}

Figure~\ref{fg:acc_results} shows the accuracy with the above configurations. PolyLUT-Add achieves the highest accuracy against all baselines on all datasets for both the linear ($D=1$) and non-linear ($D=2$) cases.

\subsubsection{Comparisons with Comparable Accuracy}
Finally, we studied area and latency with comparable accuracy. Table~\ref{tb:comparison} shows the comparisons between PolyLUT-Add and PolyLUT. $A=2$ and a lower $F$ are used for all PolyLUT-Add models (which are denoted as ``HDR-Add2'', ``JSC-XL-Add2'', ``JSC-M Lite-Add2'', see Table~\ref{tb:baselinesetup} for detailed setup). Notably, PolyLUT applied $D=4$ for HDR and JSC-XL models and $D=6$ for the JSC-M Lite model, while PolyLUT-Add used a smaller $D$. 

The PolyLUT-Add achieved a LUT reduction of 4.8$\times$, 6.5$\times$, and 13.9$\times$ for the MNIST, JCS-XL, and JSC-M Lite benchmarks, respectively. As for latency, his approach achieved a 1.6$\times$, 1.3$\times$, and 1.2$\times$ decrease for the three benchmarks, respectively. These significant reductions are attributed to a lower polynomial degree $D$ and lower $F$.

\begin{table*}[]
  \centering
  \caption{Comparison results between PolyLUT-Add and PolyLUT. The frequency and the area are collected from the Vivado post Place \& Route reports. }
  \label{tb:comparison}
  \bgroup
  \def\arraystretch{1.16}
  \setlength{\tabcolsep}{0.2cm}
  \resizebox{0.99\linewidth}{!}{ 
  \begin{tabular}{|c|lccccccc|}
  \cline{1-9}
  Dataset                        & Model                      & Accuracy$\uparrow$  & LUT    & FF     & DSP & BRAM & $F\_max$(MHz)$\uparrow$ & Latency(ns)$\downarrow$ \\ \hline\hline
  \multirow{2}{*}{MNIST}         & \bf{PolyLUT-Add (HDR-Add2, $D$=3)}              & \bf{96\%}   & \bf{14810}   &  \bf{2609}  & \bf{0}   & \bf{0}    & \bf{625}           &   \bf{10}     \\
                                 & PolyLUT (HDR, $D$=4)~\cite{polylut}        & \bf{96\%}   & 70673        & 4681        & \bf{0}   & \bf{0}    & 378           & 16       \\\hline\hline
  \multirow{2}{*}{JSC (high accuracy)}   & \bf{PolyLUT-Add (JSC-XL-Add2, $D$=3)}           & \bf{75\%}        &  \bf{36484}   &  \bf{1209}   & \bf{0}   & \bf{0}    & \bf{315}           & \bf{16}       \\
                                 & PolyLUT (JSC-XL, $D$=4)~\cite{polylut}     & \bf{75\%}        & 236541       & 2775   & \bf{0}   & \bf{0}    & 235           & 21         \\\hline\hline
  \multirow{2}{*}{JSC (low accuracy)}   & \bf{PolyLUT-Add (JSC-M Lite-Add2, $D$=3)}       & \bf{72\%}   &  \bf{895}   & \bf{189}  & \bf{0} & \bf{0}    & \bf{750}           & \bf{4}       \\
                                 & PolyLUT (JSC-M Lite, $D$=6)~\cite{polylut} & \bf{72\%}   & 12436        & 773       & \bf{0}   & \bf{0}    & 646           & 5       \\    \hline

  \end{tabular}}
  \egroup

  \begin{tablenotes}
    \footnotesize
    \item[1] \(*\): Paper reports ``LUT+FF''
\end{tablenotes}
\end{table*}
\subsection{Sparsification Evaluation}

To evaluate SparseLUT as a general approach for improving the accuracy of LUT-based DNNs, we tested its performance across several state-of-the-art methods, including LogicNets~\cite{LogicNets}, PolyLUT~\cite{polylut}, NeuraLUT~\cite{neuralut}, and our PolyLUT-Add variant enhanced with architectural optimization. Similar to the tree-structured fan-in expansion technique used in PolyLUT-Add, NeuraLUT-Assemble~\cite{neuralut-assemble} is a fan-in optimized variant of NeuraLUT, which is also evaluated in this work. We compare the effectiveness of random connectivity against the optimized connectivity provided by SparseLUT. In addition, NeuraLUT-Assemble also reports results using static sparsification, a post-training pruning strategy based on weight ranking; this static sparsification performance is also included in our evaluation.

When searching for the optimized sparse connectivity, all modes follow the same training setup to ensure consistency. The AdamW optimizer~\cite{PyTorch} in PyTorch is used, and each model is trained for 300 epochs. The threshold $T$ is set to 240 epochs, corresponding to 80\% of the total training duration. The hyperparameters $\epsilon_{1} = 10^{-12}$ and $10^{-5} \leq \epsilon_{2} \leq 10^{-4}$ are employed. The $F_i$ is set to $N$ to set the training to start from a dense model.
The primary objective is to obtain the optimized sparsity pattern $\mathcal{M}$. Full-precision training is used for faster convergence and better precision in identifying critical connections. The results presented in this section are based on these configurations, though further optimization or additional training techniques may enhance performance for specific applications.

The experiment comprises two steps: (1) derive the optimized connectivity feature mask \( \mathcal{M} \). (2) Replace the original random connectivity in the baseline models with \( \mathcal{M} \) and retrain the baselines.

\begin{table*}[]
  \centering
  \caption{Setups of baseline models.}
  \label{tb:baselinesetup}
  \bgroup
  \def\arraystretch{1.15}
  \setlength{\tabcolsep}{0.5cm}
  \resizebox{0.99\linewidth}{!}{ 
  \begin{tabular}{|c|l|lc|lcc|}
  \cline{1-7}
  \multicolumn{1}{|c|}{{Dataset}}    & \multicolumn{1}{c|}{{Method}} & \multicolumn{1}{c}{{Model}} & \multicolumn{1}{c|}{{Neurons per layer}} & \multicolumn{1}{c}{{$\beta$}} & \multicolumn{1}{c}{{$F$}} & \multicolumn{1}{c|}{{$D$}}\\ \hline \hline
  \multirow{4}{*}{MNIST}                 & PolyLUT      & HDR               & 256, 100, 100, 100, 100, 10 & 2       &  6  & 1,2  \\
                                         & PolyLUT-Add  & HDR-Add2          & 256, 100, 100, 100, 100, 10 & 2       &  4  & 1,2  \\
                                         & NeuraLUT     & HDR-5L            & 256, 100, 100, 100, 10      & 2       &  6  & ---  \\
                                         & NeuraLUT-Assemble  & HDR-5L-0.1  & 2160, 360, 2160, 360, 60, 10 & 1      &  6  & ---  \\\hline\hline
  \multirow{4}{*}{\shortstack{JSC \\ (high accuracy)}}   & PolyLUT          & JSC-XL$^1$ & 128, 64, 64, 64, 5          & 5       &  3  & 1,2     \\
                                         & PolyLUT-Add   & JSC-XL-Add2$^2$  & 128, 64, 64, 64, 5          & 5       &  2  & 1,2  \\
                                         & NeuraLUT      & JSC-5L$^1$       & 128, 128, 128, 64, 5        & 4       &  3  & ---  \\
                                         & NeuraLUT-Assemble & jsc-cernbox$^3$  & 320, 160, 80, 40, 20, 10  & 4  &  2  & ---  \\ \hline\hline
   \multirow{3}{*}{\shortstack{JSC \\ (low accuracy)}}   & PolyLUT          & JSC-M Lite      & 64, 32, 5                   & 3       &  4  & 1,2   \\
                                         & PolyLUT-Add   & JSC-M Lite-Add2  & 64, 32, 5                   & 3       &  2  & 1,2      \\
                                         & NeuraLUT      & JSC-2L           & 32, 5                       & 4       &  3  & ---   \\ \hline
   \multirow{4}{*}{CIFAR-10}         & PolyLUT      & HDR               & 256, 100, 100, 100, 100, 10 & 2       &  6  & 1,2  \\
                                         & PolyLUT-Add  & HDR-Add2          & 256, 100, 100, 100, 100, 10 & 2       &  4  & 1,2  \\
                                         & NeuraLUT     & HDR-5L            & 256, 100, 100, 100, 10      & 2       &  6  & ---  \\
                                         & NeuraLUT-Assemble  & HDR-5L-0.1  & 2160, 360, 2160, 360, 60, 10 & 1       &  6  & ---  \\\hline

  \end{tabular}}
  \egroup

  \begin{tablenotes}
    \footnotesize
    \item[1] PolyLUT with $D=1$ is equivenlent to LogicNets.
    \item[2] Remarks:  $^1$ $\beta_{i}=7$, $F_{i}$ = 2;  $^2$ $\beta_{i}=7$, $F_{i}$ = 1; $^3$ $\beta_{i}=8$, $F_{i}$ = 1;
\end{tablenotes}
\end{table*}

To ensure a fair comparison, the configurations and training hyperparameters in Step 2 are consistent with the baseline setups as reported in the original papers and summarized in Table~\ref{tb:baselinesetup}. All results in this section are obtained by retraining the models.

\begin{table}[]
\centering
  \caption{Sparsity Modes}
  \scalebox{1.0}{
  \renewcommand{\arraystretch}{1.2}
  \setlength{\tabcolsep}{0.8cm}
  \resizebox{0.95\columnwidth}{!}{ 
  \begin{tabular}{c|c}
  \hline
  \multicolumn{1}{|c|}{{Mode}}    & \multicolumn{1}{c|}{{Fixed fan-in constrain}} \\ \hline \hline
  \multicolumn{1}{|c|}{Fully Connected}  &  \multicolumn{1}{c|}{---}     \\ \hline
  \multicolumn{1}{|c|}{Random Sparsity}  &  \multicolumn{1}{c|}{\ding{51}}     \\ \hline
  \multicolumn{1}{|c|}{DeepR$^*$~\cite{deepr}}  &  \multicolumn{1}{c|}{\ding{51}}     \\ \hline
  \multicolumn{1}{|c|}{SparseLUT (ours)}  &  \multicolumn{1}{c|}{\ding{51}}     \\ \hline
\end{tabular}}}
\label{tb:4mode}
\end{table}

In Step 1, we evaluate three sparse modes with a fully connected baseline, as shown in Table~\ref{tb:4mode}:  
\begin{itemize}
    \item \textbf{Random Sparsity}: Reflects the approach currently employed by existing techniques~\cite{LogicNets,polylut,polylutadd,neuralut}.  
    \item \textbf{DeepR$^*$}: The original DeepR~\cite{deepr} cannot be directly applied to our problem. We have developed a revised version DeepR$^*$, that supports fixed fan-in, which we use as a comparison baseline.
    \item \textbf{SparseLUT}: Reflects the proposed method.
     
\end{itemize}

We then conducted a case study on the MNIST dataset, focusing on two analyses: (1) visualizing the connectivity learned by SparseLUT using heatmaps to evaluate its quality and (2) comparing test accuracy between LUT-DNN models utilizing learned versus random connectivity.

\subsubsection{Connectivity Distribution Visualization}

\begin{figure*}[t]
    \centerline{\includegraphics[width=0.9\linewidth]{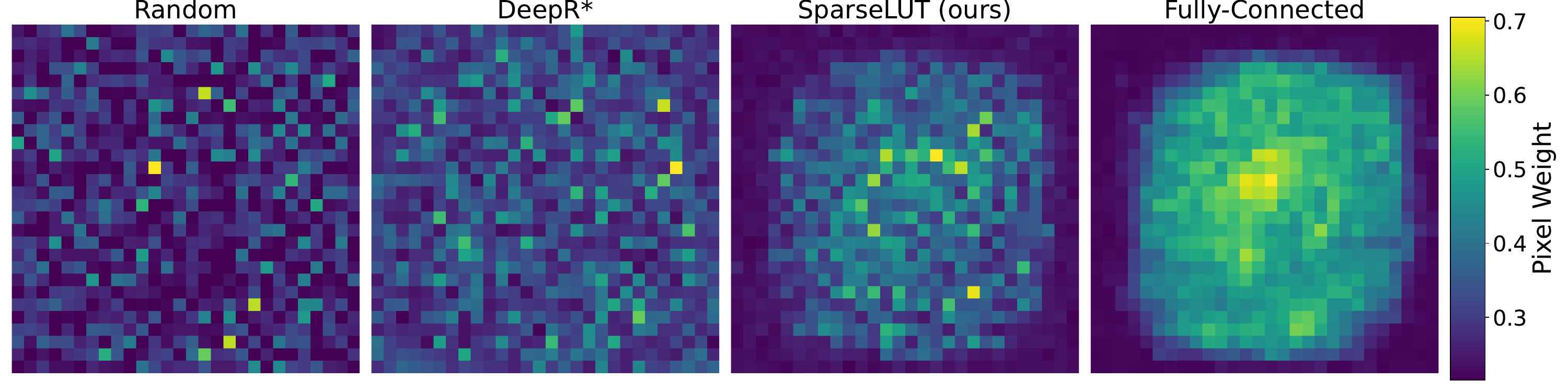}}
    \caption{Heatmaps of the average weight matrix for the first layer in three sparse modes: Random Sparsity, DeepR$^*$, and SparseLUT, with a Fully Connected mode as a baseline. } 
    \label{fg:mnist_heatmaps}
\end{figure*}

We lack ground truth to evaluate connectivity quality before testing its performance in LUT-DNNs. However, for the MNIST dataset, where handwritten digits typically appear centered~\footnote{The handwritten digits dataset visualization: \raisebox{-0.25\height}{\includegraphics[height=1.8em]{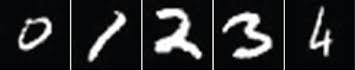}}}, it is reasonable to expect the most important connections to concentrate in the central region.

We adopted a fully connected model and three sparse models (random, DeepR$^*$, SparseLUT) with layer neurons \{256, 100, 100, 100, 10\}, consistent with the HDR-5L model from Table~\ref{tb:baselinesetup}. After training, we averaged the absolute values of the first-layer weight matrix $|W| \in \mathbb{R}^{784 \times 256}$ along the second axis and reshaped it into $W_{ave} \in \mathbb{R}^{28 \times 28}$, aligning with the input image dimensions.

Figure~\ref{fg:mnist_heatmaps} is a visualization of $W_{ave}$ as heatmaps. The fully connected model (last sub-figure) shows higher weights concentrated in the central region, aligning with our assumptions.

The first three sub-figures in Figure~\ref{fg:mnist_heatmaps} show heatmaps for sparse models, where only $F=6$ elements in each row of $|W|$ are non-zero. The random sparsity model shows a uniform distribution, reflecting unstructured connectivity. DeepR$^*$ exhibits a trend of concentrated central connectivity, suggesting meaningful adaptation to the task. SparseLUT’s heatmap more closely resembles the fully connected case, indicating that SparseLUT effectively learns an optimized connectivity pattern, prioritizing the central region in alignment with the dataset's structure.

\subsubsection{Comparison with baseline works}

\begin{figure}[t]
    \centerline{\includegraphics[width=1.0\linewidth]{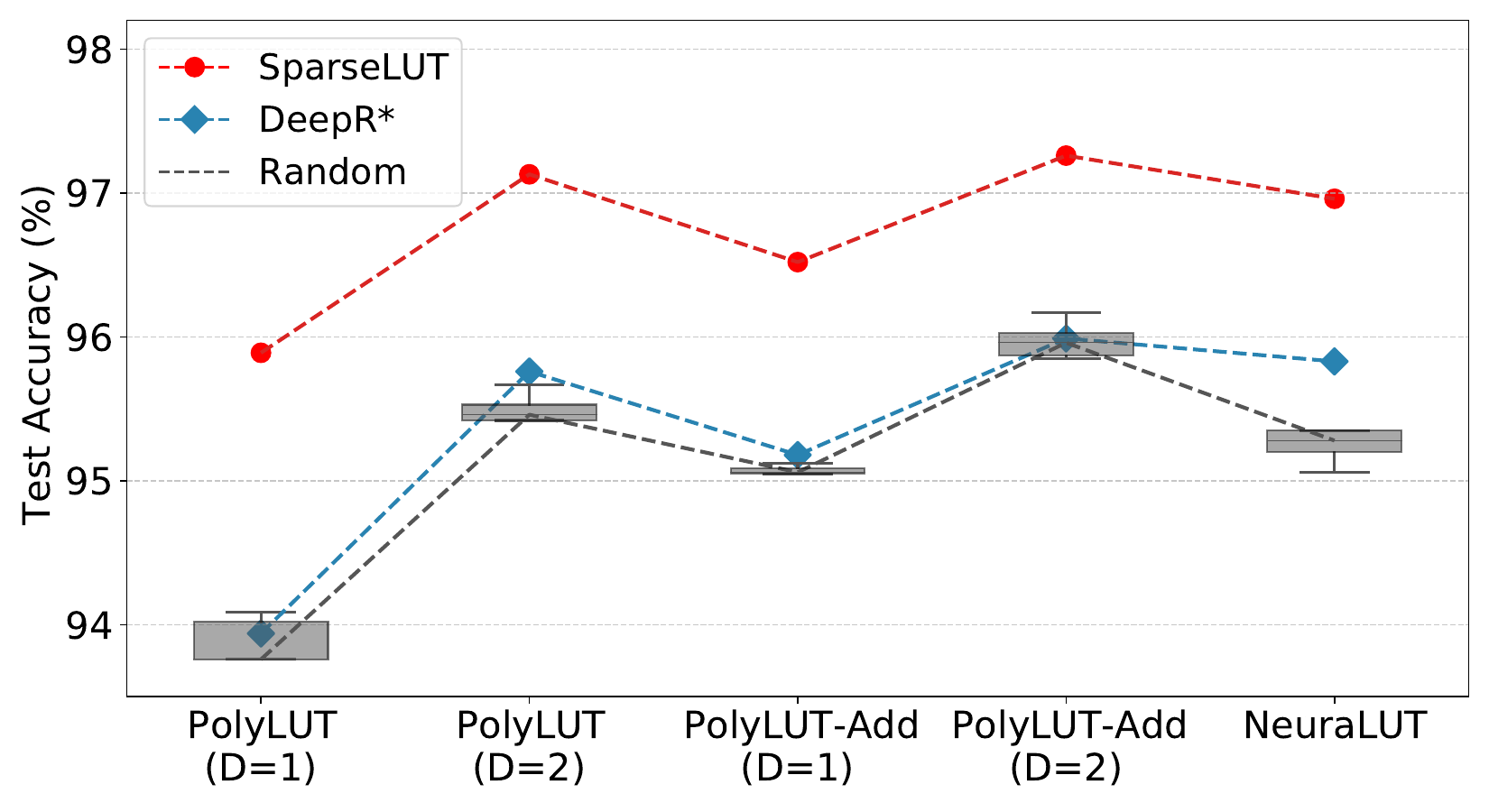}}
    \caption{Test accuracy for five models with random sparsity (5 seeds), DeepR$^*$, and SparseLUT. } 
    \label{fg:mnist_boxplot}
\end{figure}

Figure~\ref{fg:mnist_boxplot} further compares the test accuracy of applying different connectivity patterns to five models---PolyLUT ($D=1$), PolyLUT ($D=2$), PolyLUT-Add ($D=1$), PolyLUT-Add ($D=2$), and NeuraLUT---under three sparsity configurations: random sparsity (boxplots with five seeds), DeepR$^*$, and SparseLUT. For DeepR$^*$ and SparseLUT, the optimized accuracy points are plotted separately for each model.

The results show that both DeepR$^*$ and SparseLUT outperform random sparsity, but the advantages of SparseLUT are more pronounced. DeepR$^*$ achieves limited improvements over random sparsity in certain cases, such as PolyLUT ($D=1$) and PolyLUT-Add ($D=2$), where its accuracy points overlap with the box regions of random sparsity, despite being slightly above the mean line. In contrast, SparseLUT consistently delivers significant gains, exceeding the random sparsity range by 1.4\%--2.1\%. This performance gap highlights the effectiveness of SparseLUT’s non-greedy training approach, which provides a larger parameter search space in the early stages, allowing the model to explore a broader range of sparsity patterns.

\subsubsection{Test Accuracy Analysis}

\begin{table*}[]
  \centering
  \caption{Performance comparison with accuracy reported from prior works. Accuracy$^{+opt}$ indicates the accuracy with optimized connectivity by SparseLUT.}
  \label{tb:sparsecomparison}
  \bgroup
  \def\arraystretch{1.30}
  \setlength\tabcolsep{3.3mm}
  \scalebox{1.0}{
  \begin{tabular}{|c|l|l|c|l|c|c|}
  \cline{1-7}
  \multicolumn{1}{|c|}{{Dataset}}    & \multicolumn{1}{c|}{{Setup}} & \multicolumn{1}{c|}{{Model}} & \multicolumn{1}{c|}{{Accuracy}} & \multicolumn{1}{c|}{{Accuracy$^{+opt}$}} & \multicolumn{1}{c|}{{Dense}}\\ \hline \hline
  \multirow{6}{*}{MNIST}         &  HDR($D$=1)~\cite{polylut}                & PolyLUT      & 93.76\%       & \bf{95.89\%}  & \multirow{4}{*}{98.55\%}  \\ 
                    \cline{2-5}  &  HDR($D$=2)~\cite{polylut}                & PolyLUT      & 95.42\%       & \bf{97.13\%}  & \\ 
                    \cline{2-5}  &  HDR-Add2($D$=1)~\cite{polylutadd}        & PolyLUT-Add  & 95.09\%       & \bf{96.52\%}  & \\ 
                    \cline{2-5}  &  HDR-Add2($D$=2)~\cite{polylutadd}        & PolyLUT-Add  & 95.87\%       & \bf{97.26\%}  & \\ 
                    \cline{2-6}  &  HDR-5L~\cite{neuralut}                   & NeuraLUT     & 95.20\%       & \bf{96.96\%}  & 98.61\%\\ 
                    \cline{2-6}  &  HDR-5L-0.1~\cite{neuralut-assemble}  & NeuraLUT-Assemble & 95.52\%, 97.90\% & \bf{98.10\%}  & 98.87\% \\ \hline\hline
  \multirow{5}{*}{JSC (high accuracy)} & JSC-XL($D$=1)~\cite{polylut}        & PolyLUT      & 74.48\%       & \bf{74.65\%}  & \multirow{4}{*}{75.46\%} \\ 
                     \cline{2-5} &  JSC-XL($D$=2)~\cite{polylut}             & PolyLUT      & 74.94\%       & \bf{75.01\%}  &\\ 
                     \cline{2-5} &  JSC-XL-Add2($D$=1)~\cite{polylutadd}     & PolyLUT-Add  & 74.64\%       & \bf{74.74\%}  &\\ 
                     \cline{2-5} &  JSC-XL=Add2($D$=2)~\cite{polylutadd}     & PolyLUT-Add  & 74.98\%       & \bf{75.04\%}  &\\
                     \cline{2-6} &  JSC-5L~\cite{neuralut}                   & NeuraLUT     & 74.93\%       & \bf{74.98\%}  & 75.31\% \\ 
                     \cline{2-6} &  jsc-cernbox~\cite{neuralut-assemble}  & NeuraLUT-Assemble & 74.85\%, 75.00\%  & \bf{75.03\%}  & 75.26\% \\\hline\hline
   \multirow{5}{*}{JSC (low accuracy)} & JSC-M Lite($D$=1)~\cite{polylut}                   & PolyLUT      & 71.65\%       & \bf{72.10\%}  & \multirow{4}{*}{72.48\%}\\ \cline{2-5}
                                 &  JSC-M Lite($D$=2)~\cite{polylut}         & PolyLUT      & 71.98\%       & \bf{72.15\%}  &\\ \cline{2-5}
                                 &  JSC-M Lite-Add2($D$=1)~\cite{polylutadd} & PolyLUT-Add  & 71.53\%       & \bf{71.96\%}  &\\ \cline{2-5}
                                 &  JSC-M Lite-Add2($D$=2)~\cite{polylutadd} & PolyLUT-Add  & 71.90\%       & \bf{72.24\%}  &\\ \cline{7-7}
                    \cline{2-6}  &  JSC-2L~\cite{neuralut}                   & NeuraLUT     & 72.01\%       & \bf{72.95\%}  & 73.34\%  \\ \hline\hline
   \multirow{6}{*}{CIFAR-10} &  HDR($D$=1)~\cite{polylut}            & PolyLUT      &  38.83\% &  \bf{40.99\%} & \multirow{4}{*}{60.95\%}   \\ 
                    \cline{2-5}  &  HDR($D$=2)~\cite{polylut}            & PolyLUT     &  42.66\% & \bf{46.10\%} & \\ 
                    \cline{2-5}  &  HDR-Add2($D$=1)~\cite{polylutadd}    & PolyLUT-Add  &  40.81\% & \bf{42.74\%} & \\ 
                    \cline{2-5}  &  HDR-Add2($D$=2)~\cite{polylutadd}    & PolyLUT-Add  &  44.27\%  &  \bf{46.82\%} & \\ 
                    \cline{2-6}  &  HDR-5L~\cite{neuralut}               & NeuraLUT     & 42.98\% &  \bf{46.38\%} & 61.38\% \\ 
                    \cline{2-6}  &  HDR-5L-0.1~\cite{neuralut-assemble}  & NeuraLUT-Assemble & 44.84\% &  \bf{46.27\%}  &  62.51\%\\ \hline

  \end{tabular}}
  \egroup

  \begin{tablenotes}
    \footnotesize
    \item[1] LogicNets~\cite{LogicNets} cases are equivalent to all baselines with $D=1$.
\end{tablenotes}
\end{table*}

\begin{figure}[]
    \centerline{\includegraphics[width=1.0\linewidth]{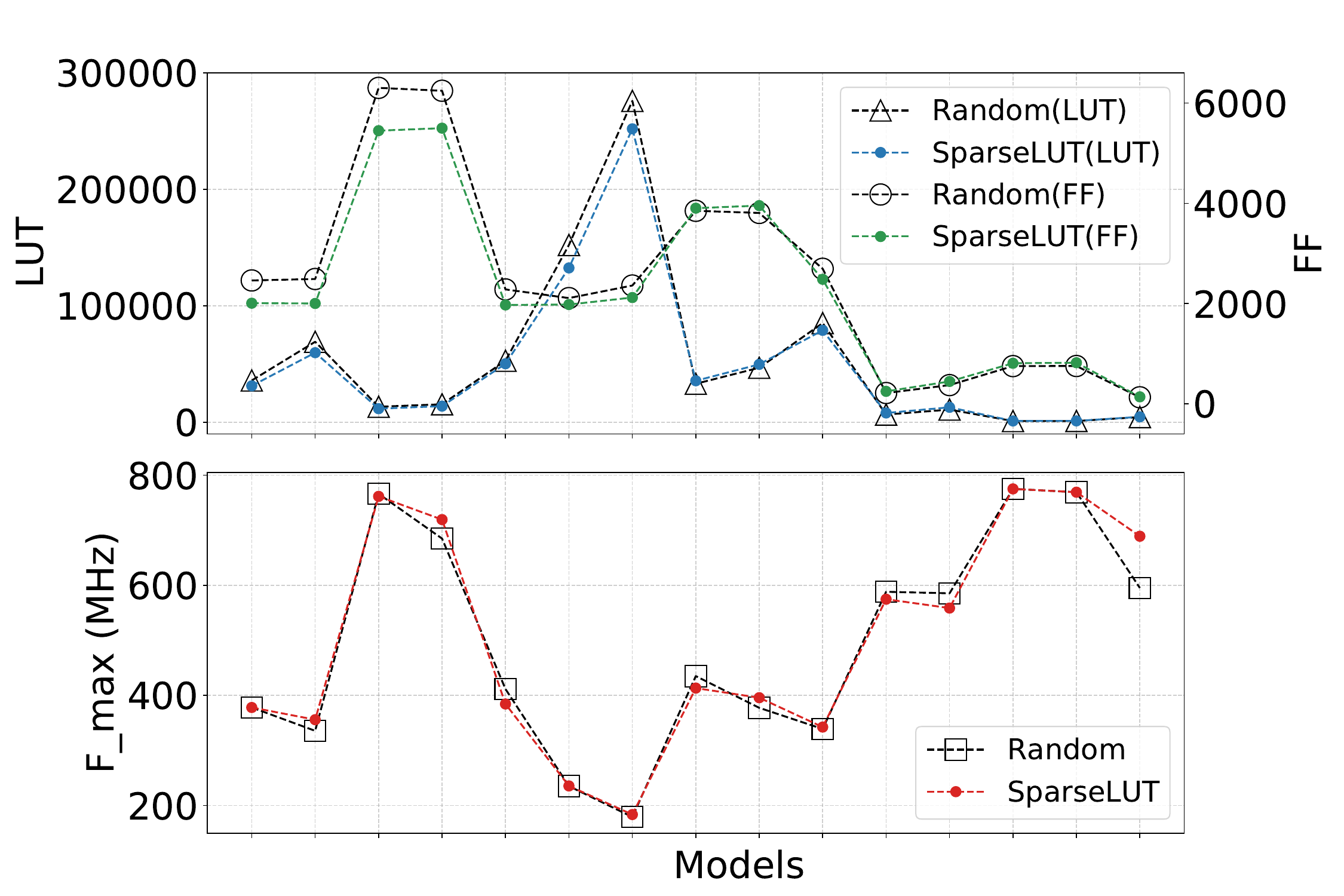}}
    \caption{Post place \& route hardware comparison between models (in the same order from left to right (PolyLUT to NeuraLUT) as Table~\ref{tb:sparsecomparison}) with/without connectivity from SparseLUT.} 
    \label{fg:lutff}
\end{figure}

In this subsection, we present the performance of SparseLUT across multiple baseline models. For the MNIST and JSC datasets, accuracy values under random sparsity are obtained from referenced papers and open-source implementations. As shown in Table~\ref{tb:sparsecomparison}, SparseLUT consistently outperforms random sparsity configurations across all baseline models, as well as the static-sparsification-enhanced accuracy of the NeuraLUT-Assemble model (with random accuracy reported on the left and optimized accuracy on the right).

To further assess the generalizability of the proposed connectivity optimization on model accuracy, results on the CIFAR-10 dataset are reported in Table~\ref{tb:sparsecomparison}. The comparison demonstrates the superiority of SparseLUT over random sparsity across the evaluated models. The relatively low accuracy achieved on CIFAR-10 can be attributed to the simple architectural design of existing LUT-based DNNs. Developing LUT-based DNNs that achieve practical accuracy (e.g., exceeding 90\%) on complex tasks such as CIFAR-10, or even larger benchmarks, remains an open challenge in the FPGA community.  

To assess the ceiling for accuracy improvement achievable through optimized sparsity, we use the fully connected model of LogicNets ({\em i.e.}, polynomial degree $D=1$) as a fair comparison point for PolyLUT ($D=1$) and PolyLUT-Add ($D=1$) configurations. For models with $D>2$ or NeuraLUT's network-in-network architecture, fully connected implementations result in an exponential increase in parameters, rendering such implementations impractical. Consequently, the dense accuracy shown in the last column of Table~\ref{tb:sparsecomparison} for these configurations is considered a reasonable indicator rather than an upper bound.

The results highlight notable differences in accuracy deltas ($\delta$ = \textit{Dense Accuracy} - \textit{Random Sparsity Accuracy}) across datasets and models. For MNIST and CIFAR-10, SparseLUT achieves a substantial accuracy boost, attributed to the large $\delta$ exceeding 3\%. In contrast, for most models in the JSC dataset, this $\delta$ is much smaller—approximately 0.8\%—explaining the comparatively modest accuracy improvements achieved with optimized connectivity. A notable exception within the JSC dataset is the JSC-2L model on NeuraLUT (last row), which exhibits a $\delta$ of 1.33\%. SparseLUT significantly boosts its accuracy from 72.01\% to 72.95\%, likely due to the higher word length ($\beta = 4$) of the JSC-2L model compared to the 3-bit word length ($\beta = 3$) in other JSC-M Lite models.

These findings reveal that SparseLUT’s effectiveness is closely tied to the connectivity optimization potential of the underlying model, as reflected in the $\delta$ between dense and sparse configurations. This highlights the critical role of dataset characteristics and model architecture in determining the extent of accuracy improvements achievable with SparseLUT.

SparseLUT does not alter the number of LUT entries in the generated RTL design; Figure~\ref {fg:lutff} illustrates the post-Place\&Route results for area (LUTs and flip-flops (FFs)) and maximum frequency (F\_max) from Vivado, constrained within the same clock cycle. Notably, SparseLUT achieves similar hardware consumption, and the F\_max comparison shows no speed penalty. Therefore, we conclude that the accuracy improvements achieved by SparseLUT are realized without incurring additional hardware and latency overhead.

Table~\ref{tb:comparison_all} compares the connectivity-optimized models discussed above with other low-latency machine learning methods. Since DWN achieves 57.42\% accuracy on CIFAR-10, whereas all traditional LUT-based DNNs remain below 50\% accuracy (see Table~\ref{tb:sparsecomparison}), evaluating their hardware cost at this stage is not meaningful. Consequently, exploring the potential of both approaches remains an open direction for future research within the FPGA community.

\begin{table*}[]
  \centering
  \caption{Comparison of generalized LUT-DNN against other ultra-low latency machine learning techniques.}
  \label{tb:comparison_all}
  \bgroup
  \def\arraystretch{1.2}
  \setlength{\tabcolsep}{0.05cm}
  \resizebox{0.99\linewidth}{!}{ 
  \begin{tabular}{|c|lcccccccc|}
  \cline{1-10}
  Dataset                        & Model                  & Accuracy$\uparrow$    & Accuracy$^{+opt}\uparrow$  & LUT    & FF     & DSP & BRAM & $F\_max$(MHz)$\uparrow$ & Latency(ns)$\downarrow$ \\ \hline\hline
  \multirow{10}{*}{MNIST}  
                & PolyLUT (HDR, $D$=2)~\cite{polylut}                      & 95.42\%     & 97.13\%     & 59790      & 2003      & \bf{0}   & \bf{0}   & 356      & 17       \\ 
                & PolyLUT-Add (HDR-Add2, $D$=2)~\cite{polylutadd}          & 95.87\%     & 97.26\%     & 13890      & 2768      & \bf{0}   & \bf{0}   & 635      & 8        \\
                & NeuraLUT (HDR-5L)~\cite{neuralut}                        & 95.20\%     & 96.96\%     & 50271      & 1973      & \bf{0}   & \bf{0}   & 384      & 13       \\
                & NeuraLUT-Assemble (HDR-5L-0.1)~\cite{neuralut-assemble}  & 97.90\%     &\bf{98.10\%} & 5076       & 725       & \bf{0}   & \bf{0}   & 863      & \bf{2.1} \\
   \cline{2-10} & DWN~\cite{dwn}                                                    & 97.8\%      & -           & \bf{2092}  & 1757      & \bf{0}   & \bf{0}   & \bf{873} & 9.2      \\
                & DWN$^{arxiv}$~\cite{dwn2}                                         & \bf{98.3\%} & -           & 4082       & 3385      & \bf{0}   & \bf{0}   & 827      & 6        \\
                & TreeLUT~\cite{treelut}                                            & 96.6\%      & -           & 4478       & \bf{597}  & \bf{0}   & \bf{0}   & 791      & 2.5      \\
                & PolyLUT$^{extension}$~\cite{polylut2}                             & 97.5\%      & -           & 75131      &  4668     & \bf{0}   & \bf{0}   & 353      & 17       \\
                & FINN~\cite{finn}                                                  & 96\%        & -           & 91131      &  -        & \bf{0}   & 5        & 200      & 310      \\
                & \texttt{hls4ml}~\cite{hls4ml}                                     & 95\%        & -           & 260092     & 165513    & \bf{0}   & 345      & 200      & 190      \\\hline\hline
  \multirow{9}{*}{JSC (high accuracy)}  
                & PolyLUT (JSC-XL, $D$=2)~\cite{polylut}                   & 74.94\%     & 75.01\%     & 252077   & 2121    & \bf{0}  & \bf{0}    & 183          & 27        \\
                & PolyLUT-Add (JSC-XL-Add2, $D$=2)~\cite{polylutadd}       & 74.98\%     &\bf{75.04\%} & 49766    & 1976    & \bf{0}  & \bf{0}    & 395          & 13        \\
                & NeuraLUT (JSC-5L)~\cite{neuralut}                        & 74.93\%     & 74.98\%     & 79012    & 2485    & \bf{0}  & \bf{0}    & 342          & 15        \\
                & NeuraLUT-Assemble (jsc-cernbox)~\cite{neuralut-assemble} & 75.00\%     & 75.03\%     & 8556     & 1335    & \bf{0}  & \bf{0}    & 352          & 5.7       \\
 \cline{2-10}   & Duarte {\em et al.}~\cite{duarte2018fast}                         & 75\%        & -           & \multicolumn{2}{c}{88797$^*$} & 954  & \bf{0} & 200     & 75        \\
                & TreeLUT~\cite{treelut}                                            & 75.6\%      & -           &\bf{2234} &\bf{347} & \bf{0}  & \bf{0}    &  735         & \bf{2.7}  \\
                & DWN$^{arxiv}$~\cite{dwn2}                                         & \bf{76.3\%} & -           & 4972     & 3305    & \bf{0}  & \bf{0}    & \bf{827}     & 7.3  \\
                & PolyLUT$^{extension}$~\cite{polylut2}                             & 75.1\%      & -           & 246071   & 12384   & \bf{0}  & \bf{0}   &  203          & 25        \\
                & Fahim {\em et al.}~\cite{fahim2021hls4ml}                         & 76.2\%      & -           & 63251    & 4394    & 38      & \bf{0}    & 200          & 45 \\\hline\hline
  \multirow{6}{*}{JSC (low accuracy)} 
                & LogicNets~\cite{LogicNets}                               & 71.65\%     & 72.10\%      & 7995      & 249       & \bf{0}   & \bf{0}    & 574       & 5         \\ 
                & PolyLUT (JSC-M Lite, $D$=2)~\cite{polylut}               & 71.98\%     & 72.15\%      & 12855     & 447       & \bf{0}   & \bf{0}    & 558       & 5         \\
                & PolyLUT-Add (JSC-M Lite-Add2, $D$=2)~\cite{polylutadd}   & 71.90\%     & 72.24\%      & 1197      & 410       & \bf{0}   & \bf{0}    & 769       & 4         \\
                & NeuraLUT (JSC-2L)~\cite{neuralut}                        & 72.01\%     & \bf{72.95\%} & 4679      & 140       & \bf{0}   & \bf{0}    & 689       & 3         \\
 \cline{2-10}   & DWN$^{arxiv}$~\cite{dwn2}                                         & \bf{74.0\%} & -            & \bf{110}  & \bf{72}   & \bf{0}   & \bf{0}    & \bf{1094} & \bf{1.5}  \\
                & PolyLUT$^{extension}$~\cite{polylut2}                             & 72.5\%      & -            & 10169     & 631       & \bf{0}  & \bf{0}     & 598       & 5        \\
                
                                 \hline

  \end{tabular}}
  \egroup

  \begin{tablenotes}
    \footnotesize
    \item[1] 1. \(*\): Paper reports ``LUT+FF''
    \item[1] 2. DWN$^{arxiv}$ is the arXiv version of DWN with updated results, PolyLUT$^{extension}$ is the journal extension version of PolyLUT with updated results.
\end{tablenotes}
\end{table*}

\subsubsection{Run Time Results}

\begin{table}[h]
\centering
  \caption{Task Execution Time}
  \scalebox{1.0}{
  \renewcommand{\arraystretch}{1.23}
  \setlength{\tabcolsep}{1cm}
  \resizebox{0.95\columnwidth}{!}{ 
  \begin{tabular}{c|c}
  \hline
  \multicolumn{1}{|c|}{{Task}}    & \multicolumn{1}{c|}{{Time}} \\ \hline \hline
  \multicolumn{1}{|c|}{Connectivity Searching}  &  \multicolumn{1}{c|}{2.6h}     \\ \hline
  \multicolumn{1}{|c|}{LUT-DNN Training (PolyLUT)}  &  \multicolumn{1}{c|}{3.9h}     \\ \hline
  \multicolumn{1}{|c|}{RTL Generation}  &  \multicolumn{1}{c|}{1.4h}     \\ \hline
  \multicolumn{1}{|c|}{Synthesis \& Place \& Route}  &  \multicolumn{1}{c|}{2.5h}     \\ \hline
\end{tabular}}}
\label{tb:runtime}
\end{table}

Finally, Table~\ref{tb:runtime} breaks down the execution times for the various tasks involved and shows that the runtime of connectivity searching does not significantly impact the overall system performance. To illustrate this, we measured the runtime for implementing a PolyLUT ($D=2$) model on the MNIST task. Synthesis, Place, and Route were performed on a machine with an Intel Xeon E-2356G CPU and 94GB of RAM, while all other processes were executed on a system with an Intel Core i7-10700F CPU, 64GB of RAM, and an NVIDIA RTX 2060 Super GPU.

\section{Conclusion}
\label{se:Conclusion}

This paper introduced SparseLUT, a comprehensive framework that enhances the deployment of LUT-based deep neural networks (DNNs) on FPGAs through two orthogonal optimizations: architectural enhancement and training-based connectivity optimization.

At the architectural level, we proposed an efficient structure that combines multiple PolyLUT sub-neurons via an adder, significantly reducing hardware cost and latency without sacrificing accuracy. Experimental results show that this approach achieves 2.0$\times$–13.9$\times$ LUT savings and 1.2$\times$–1.6$\times$ latency reduction across diverse benchmarks such as MNIST and Jet Substructure Classification.

On top of this, we introduced a non-greedy dynamic sparsity training algorithm that optimizes the connectivity pattern of each neuron under fixed fan-in constraints, with no additional LUT or routing overhead. By intelligently pruning and regrowing connections, SparseLUT consistently improves accuracy over existing LUT-DNNs such as LogicNets~\cite{LogicNets}, PolyLUT~\cite{polylut}, and NeuraLUT~\cite{neuralut}, achieving up to 2.13\% higher accuracy on MNIST and 0.94\% on Jet Substructure Classification.

\newpage
\bibliographystyle{unsrt}
\bibliography{ref}

\end{document}